\documentclass[trackchanges,twocolumn]{aastex701}

\begin{document}

\title{Discovery of a New Electron Spectrum Produced by Fast Cooling of Relativistic Electrons}

\author{Z. Lucas Uhm}
\affiliation{Korea Astronomy and Space Science Institute, Daejeon 34055, Republic of Korea; z.lucas.uhm@gmail.com}
\email{z.lucas.uhm@gmail.com}


%% AAS journals all have a 250 word limit for the abstract. 

\begin{abstract}
In energetic astrophysical sources like the gamma-ray bursts (GRBs) and blazars, the relativistic electrons lose their energy efficiently through synchrotron radiation and inverse Compton (IC) scattering and can rapidly cool to energies below the minimum acceleration energy $\gamma_m$. It is widely known that such electrons develop a power-law spectrum $\gamma_e^{-2}$ below $\gamma_m$. However, this power-law spectrum is valid only if a steady-state population of electrons is assumed. In general, the cooled electrons below $\gamma_m$ evolve in a non-steady state and their spectrum deviates significantly from the power-law $\gamma_e^{-2}$. We present here a new electron spectrum $f_{\rm U}$ that describes this fast-cooled, generally-evolving, and non-steady-state population of electrons below $\gamma_m$. We consider two cooling cases: (1) when the electrons undergo radiative (i.e., synchrotron + IC) cooling, and (2) when the electrons experience both radiative and adiabatic cooling. We first derive $f_{\rm U}$ analytically and find an explicit expression of $f_{\rm U}$ for each cooling case. Secondly, we show that $f_{\rm U}$ is a solution to a Fokker-Planck equation in the entire energy range below $\gamma_m$ and derive a general version of $f_{\rm U}$ for each cooling case. We also compare $f_{\rm U}$ to a detailed numerical calculation of a blazar electron spectrum and find an excellent agreement for the whole energy range below $\gamma_m$. Thus, $f_{\rm U}$ gives us a general electron spectrum to describe the entire cooled electrons below $\gamma_m$. Our discovery of a new fast-cooling solution $f_{\rm U}$ below $\gamma_m$ may find it applicable in a wide range of energetic phenomena in high-energy astrophysics.
\end{abstract}



\section{Introduction}

Both powered by relativistic jets emerging from central black holes \citep{nemmen2012} and beamed toward the Earth \citep{madau1987,kumar2015}, the gamma-ray bursts (GRBs) and blazars are among the most energetic astrophysical phenomena in the Universe. While GRBs are short-lived flashes arising from collapsing stars \citep{woosley1993,macfadyen1999} or neutron star mergers \citep{eichler1989,lyman2018}, the blazars are persistent lighthouses driven by active galactic nuclei (AGN) at the center of galaxies \citep{antonucci1993,marscher2008}. 

Although they differ tremendously in their origin, duration, and physical scale, they both launch plasma outflows propagating nearly at the speed of light. In the comoving frame of these relativistic jets, the electrons are accelerated to ultra-relativistic energies via shocks \citep[e.g.,][]{spitkovsky2008} or magnetic reconnection events \citep[e.g.,][]{zhang_hao2023} and radiate their energy across the electromagnetic spectrum powering the bright emission of GRBs \citep{rees1994,meszaros1997,sari1998} and blazars \citep{marscher1985,kirk1998,bottcher2007}.

In such energetic astrophysical sources, the accelerated electrons lose their energy efficiently through synchrotron radiation and inverse Compton (IC) scattering \citep{rybicki1979} and can rapidly cool to lower energies below the minimum acceleration energy. It is widely known that the relativistic electrons undergoing such radiative losses develop the following spectrum in this fast-cooling regime \citep{sari1998,granot2002}
\begin{equation}
\label{eq:fge}
f(\gamma_e) \equiv \frac{{\rm d} N_e}{{\rm d} \gamma_e} \propto \gamma_e^{-2} 
\quad \quad  \mbox{for} \quad \gamma_e < \gamma_m,
\end{equation}
where $N_e$ is the number of electrons, $\gamma_e$ is the Lorentz factor of electrons, and $\gamma_m$ is the minimum injection Lorentz factor of accelerated electrons.

The power-law electron spectrum in Equation (\ref{eq:fge}) is valid only if a steady-state population of electrons is assumed. This can be shown using a Fokker-Planck equation,
\begin{equation}
\label{eq:fokker_planck}
\frac{\partial}{\partial t} \left[ f(\gamma_e,t) \right] + 
\frac{\partial}{\partial \gamma_e} \left[ \dot \gamma_e\, f(\gamma_e,t) \right] = Q(\gamma_e,t),
\end{equation}
which governs how an electron spectrum $f(\gamma_e,t)$ should evolve in time and energy space. Here, $t$ is the time measured in the comoving frame of the electron population\footnote{Note that we do not add a prime to the comoving time $t$ in this work, for convenience.}, $\dot \gamma_e = {\rm d} \gamma_e/ {\rm d} t$ describes cooling of electrons, and $Q$ is a source function above $\gamma_m$ such that $Q = 0$ for $\gamma_e < \gamma_m$. If a steady-state is enforced, i.e., $\partial [f(\gamma_e, t)]/\partial t = 0$, Equation (\ref{eq:fokker_planck}) yields $f(\gamma_e) \propto \dot \gamma_e^{-1} \propto \gamma_e^{-2}$ below $\gamma_m$ for radiative cooling. 

However, there is no reason for the electron population to remain in such a steady-state, and in general, the electron population evolves into a non-steady state. Indeed, \cite{uhm2014} numerically showed that the power-law spectrum in Equation (\ref{eq:fge}) is achieved only for a steady state, and in general the electron spectrum below $\gamma_m$ evolves in a non-steady state and deviates significantly from Equation (\ref{eq:fge}).

Therefore, a more general electron spectrum describing a non-steady state population of electrons is called for.

\section{New Electron Spectrum}

We present here a new electron spectrum $f_{\rm U}$ produced by cooling of relativistic electrons in the fast-cooling regime. We show that the spectrum $f_{\rm U}$ provides us with a general electron spectrum to describe the cooled population of electrons below $\gamma_m$, which is in a non-steady state. We show that $f_{\rm U}$ is a solution to the Fokker-Planck equation above and gives us a new fast-cooling solution below $\gamma_m$. It takes a simple form
\begin{equation}
\label{eq:fge_U}
f_{\rm U} (\gamma_e, t)  =
\frac{A(t)}{\gamma_e^2 \left(1+\gamma_0 (t)/\gamma_e \right)^{\beta}},
\end{equation}
where $A(t)$ shows an overall magnitude, $\gamma_0 (t)$ gives a characteristic Lorentz factor of the spectrum, and an exponent $\beta \, (>0)$ determines its asymptotic behavior at low energies. It asymptotically leads to
\begin{eqnarray}
\label{eq:fge_U_asym}
f_{\rm U} (\gamma_e) \propto
\left\{ 
\begin{array}{ll}
\gamma_e^{-2}        \quad \quad & \mbox{for} \quad \gamma_e \gg \gamma_0, \\
\gamma_e^{-2+\beta}  \quad \quad & \mbox{for} \quad \gamma_e \ll \gamma_0.
\end{array}
\right.
\end{eqnarray}
We show below that $f_{\rm U}$ in Equation (\ref{eq:fge_U}) gives a general electron spectrum below $\gamma_m$ for two cooling cases: (1) when the electrons undergo radiative cooling, and (2) when the electrons experience both radiative and adiabatic cooling.

\subsection{Analytical Derivation of $f_{\rm U}$} \label{section:anal_deriv}

We first present an analytical approach to derive the electron spectrum $f_{\rm U}$ in Equation (\ref{eq:fge_U}). We consider a region of electrons and allow that fresh, accelerated electrons are continuously injected into the region at an injection rate
\begin{equation}
R_{\rm inj} \equiv \frac{{\rm d} N_e}{{\rm d} t}.
\end{equation} 
The source function $Q(\gamma_e,t)=0$ for $\gamma_e < \gamma_m$ while $Q(\gamma_e,t)$ is non-zero for $\gamma_e \geq \gamma_m$, such that
\begin{equation}
R_{\rm inj} = \int_{\gamma_m}^\infty Q(\gamma_e,t) {\rm d} \gamma_e. 
\end{equation}

\subsubsection{Radiative Cooling}

The radiative cooling of an electron with the Lorentz factor $\gamma_e$ experiencing both synchrotron and IC losses is given by
\begin{equation}
\label{eq:ge_dot_rad}
\dot \gamma_e = -g(t)\, \gamma_e^2,
\end{equation}
with
\begin{equation}
\label{eq:g_t}
g(t) \equiv \frac{\sigma_T}{6 \pi m_e c}\, B^2\, (1+Y),
\end{equation}
where $\sigma_T$ is the Thomson cross section, $m_e$ is the electron mass, $c$ is the speed of light, and $B$ is the magnetic field strength measured in the comoving frame. The IC cooling is taken into account in the Thomson regime employing the IC parameter $Y = P_{\rm IC}/P_{\rm syn}$, where $P_{\rm IC}$ and $P_{\rm syn}$ are the IC and synchrotron emission power of an electron with $\gamma_e$, respectively.

For this part of analytical derivation, we consider a power-law dependence of the magnetic field strength $B$ and the IC parameter $Y$ as follows
\begin{equation}
\label{eq:B_Y_powerlaw}
B = B_0\, (r/r_0)^{-b}
\quad \mbox{and} \quad 
Y = Y_0\, (r/r_0)^y,
\end{equation}
where $r$ is the distance measured from the jet base or the central source to the region of the electron population, $B_0$ and $Y_0$ are a normalization constant at distance $r_0$, respectively, and $b$ and $y$ are a power-law index, respectively. The comoving time $t$ is given by $t = r/(c \Gamma)$ for a constant bulk motion of the electron region with the Lorentz factor $\Gamma$ (when $\Gamma \gg 1$). Equation (\ref{eq:g_t}) yields then
\begin{equation}
\label{eq:g_t_powerlaw}
g(t) = g_0\, t^{-2b+y},
\end{equation}
with
\begin{equation}
\label{eq:g_0}
g_0 \equiv \frac{\sigma_T}{6 \pi m_e c}\, B_0^2\, Y_0\, (c \Gamma/r_0)^{-2b+y},
\end{equation}
where $Y \gg 1$ is assumed. 
Also, we consider a power-law dependence for the injection rate as follows
\begin{equation}
R_{\rm inj} = R_{\rm inj}^0\, (r/r_0)^q,
\end{equation}
where $R_{\rm inj}^0$ is a normalization constant at distance $r_0$, and $q$ is a power-law index. Since $t = r/(c \Gamma)$, we have
\begin{equation}
\label{eq:h_0}
R_{\rm inj} = h_0\, t^q
\quad \mbox{with} \quad 
h_0 \equiv R_{\rm inj}^0\, (c \Gamma/r_0)^q.
\end{equation}
The constants $g_0$ and $h_0$ in Equations (\ref{eq:g_0}) and (\ref{eq:h_0}) give us a measure of the electron cooling rate and the injection rate, respectively.

Now dividing Equation (\ref{eq:ge_dot_rad}) by $-\gamma_e^2$, we have
\begin{equation}
\label{eq:1_ge_rad}
\frac{\rm d}{{\rm d} t} \left( \frac{1}{\gamma_e} \right) =
g(t) = g_0\, t^{-2b+y}.
\end{equation}
Integrating Equation (\ref{eq:1_ge_rad}) from $t_i$ to $t_f$, we get
\begin{equation}
\label{eq:ge_f_rad}
\frac{1}{\gamma_{e, f}} = 
\frac{1}{\gamma_{e,i}} + \frac{g_0}{1-2b+y} \left[ t_f^{1-2b+y} - t_i^{1-2b+y} \right],
\end{equation}
where $\gamma_{e,i}$ and $\gamma_{e,f}$ denote for $\gamma_e (t_i)$ and $\gamma_e (t_f)$, respectively. We consider an infinitesimal group of $\delta N_e$ electrons injected at time $t_i$ during an interval [$t_i$, $t_i + \delta t_i$]. This injection occurs only above $\gamma_m$ (i.e., $\gamma_{e,i} \geq \gamma_m$), and its injection rate is given by
\begin{equation}
\label{eq:R_inj_h0}
R_{\rm inj} = \frac{\delta N_e}{\delta t_i} = h_0\, t_i^q.
\end{equation}
In the fast-cooling regime, these electrons cool rapidly such that $\gamma_{e,f} \ll \gamma_{e,i}$. Equation (\ref{eq:ge_f_rad}) yields then
\begin{equation}
\label{eq:ge_f_rad2}
\frac{1}{\gamma_{e, f}} = 
\frac{g_0}{1-2b+y} \left[ t_f^{1-2b+y} - t_i^{1-2b+y} \right].
\end{equation}

In order to find the electron spectrum at time $t_f$, we take a variation on both sides of Equation (\ref{eq:ge_f_rad2}) at the fixed time $t_f$, which gives
\begin{equation}
\label{eq:delta_ge_f_rad}
\frac{\delta \gamma_{e, f}}{[\gamma_{e, f}]^2} = 
g_0\, t_i^{-2b+y}\, \delta t_i.
\end{equation}
Equation (\ref{eq:delta_ge_f_rad}) shows that a variation in $\gamma_{e, f}$ is induced by a variation in $t_i$, meaning that those $\delta N_e$ electrons injected at time $t_i$ during [$t_i$, $t_i + \delta t_i$] would occupy an energy interval [$\gamma_{e, f}$, $\gamma_{e, f} + \delta \gamma_{e, f}$] of the electron spectrum at time $t_f$. Therefore, combining Equations (\ref{eq:R_inj_h0}) and (\ref{eq:delta_ge_f_rad}) (by eliminating $\delta t_i$), we obtain
\begin{equation}
\label{eq:Ne_ge_f_rad}
\frac{\delta N_e}{\delta \gamma_{e, f}} =
\frac{h_0}{g_0}\,[\gamma_{e, f}]^{-2}\, t_i^{2b-y+q}.
\end{equation}
Finally, solving for $t_i$ in Equation (\ref{eq:ge_f_rad2}) and substituting it into Equation (\ref{eq:Ne_ge_f_rad}), we find the electron spectrum
\begin{equation}
\label{eq:fU_anal_rad}
\frac{\delta N_e}{\delta \gamma_e} 
= \frac{\frac{h_0}{g_0}\, t^{2b-y+q}}
{\gamma_e^2\, \left[ 1+ \left(\frac{2b-y-1}{g_0}\, t^{2b-y-1}\right) \frac{1}{\gamma_e} \right]^{\frac{2b-y+q}{2b-y-1}}},
\end{equation}
where, for clarity, we have now removed {\it f} from $t_f$ and $\gamma_{e,f}$. Equation (\ref{eq:fU_anal_rad}) is identical to $f_{\rm U} (\gamma_e, t)$ in Equation (\ref{eq:fge_U}) when
\begin{eqnarray}
\label{eq:A_anal_rad}
A(t) =&& \frac{h_0}{g_0}\, t^{2b-y+q}, \\
\label{eq:g0_anal_rad}
\gamma_0(t) =&& \frac{2b-y-1}{g_0}\, t^{2b-y-1}, \\
\label{eq:beta_anal_rad}
\beta =&& \frac{2b-y+q}{2b-y-1}.
\end{eqnarray}
Therefore, we have analytically derived the electron spectrum $f_{\rm U}$ for radiative cooling of electrons. 

The ratio $h_0/g_0$ of the injection measure $h_0$ to the cooling measure $g_0$ in Equation (\ref{eq:A_anal_rad}) shows that the overall magnitude $A$ of the electron spectrum is determined through a competition between the cooling and the injection: i.e., a strong cooling lowers the electron spectrum while a high injection boosts it up, as expected.

Equation (\ref{eq:g0_anal_rad}) shows that the characteristic Lorentz factor $\gamma_0$ is not affected by the injection: it does not include the injection measure $h_0$ or the injection index $q$ so that it completely decouples from the injection. It is inversely proportional to the cooling measure $g_0$ so that the higher the cooling is, the lower it is. Physically, $\gamma_0$ corresponds to a ``cooling Lorentz factor'', which is usually referred to as $\gamma_c$ in the field. It is instructive to show that for synchrotron cooling (i.e., $y=0$ and $Y_0=1$)\footnote{In Equation (\ref{eq:g_t_powerlaw}), we replaced $1+Y$ by $Y=Y_0\, (r/r_0)^y$, assuming $Y \gg 1$. Therefore, by setting $y=0$ and $Y_0=1$, we can obtain the case of $Y=0$ (i.e., no IC cooling), so that the radiative cooling includes synchrotron cooling only.}, Equation (\ref{eq:g0_anal_rad}) becomes
\begin{equation}
\label{eq:g0_gc}
\gamma_0(t) = (2b-1)\, \frac{6 \pi m_e c}{\sigma_T B^2 t},
\end{equation}
where we have used the relation $r=c \Gamma t$ and Equations (\ref{eq:B_Y_powerlaw}) and (\ref{eq:g_0}). Equation (\ref{eq:g0_gc}) is nearly the same as the commonly-known $\gamma_c$ given in \cite{sari1998}: it differs by a factor of $2b-1$. 

Equation (\ref{eq:beta_anal_rad}) shows that the cooling and injection indices both enter into the expression of $\beta$. Hence, the low-energy behavior of the electron spectrum is influenced by the cooling and injection together. In a special case of $q=0$ (with a constant injection rate), the exponent $\beta$ is decided by cooling only.

\subsubsection{Radiative and Adiabatic Cooling}

We now consider both radiative and adiabatic cooling that an electron with $\gamma_e$ experiences as follows
\begin{equation}
\label{eq:ge_dot_rad_adia}
\dot \gamma_e = -g(t)\, \gamma_e^2 - \frac{2}{3 t} \gamma_e,
\end{equation}
where the second term on the right-hand side describes the adiabatic cooling for a conical flow expanding radially with a constant bulk Lorentz factor.
By dividing Equation (\ref{eq:ge_dot_rad_adia}) by $-\gamma_e^2$, we get \citep{uhm2012,uhm2014}
\begin{equation}
\label{eq:1_ge_rad_adia}
\frac{\rm d}{{\rm d} t} \left( \frac{1}{\gamma_e} \right) =
g_0\, t^{-2b+y} + \frac{2}{3t} \left(\frac{1}{\gamma_e} \right).
\end{equation}
Equation (\ref{eq:1_ge_rad_adia}) has a general solution
\begin{equation}
\label{eq:1_ge_rad_adia2}
\frac{1}{\gamma_e} = 
t^{2/3} \left[ C + \frac{g_0}{1/3-2b+y}\, t^{1/3-2b+y} \right],
\end{equation}
where $C$ is an integration constant. Evaluating Equation (\ref{eq:1_ge_rad_adia2}) at $t_i$ and $t_f$ and subtracting one result from the other, we obtain
\begin{equation}
\label{eq:ge_f_rad_adia}
\frac{t_f^{-2/3}}{\gamma_{e, f}} = 
\frac{t_i^{-2/3}}{\gamma_{e, i}} + \frac{g_0}{1/3-2b+y} \left[ t_f^{1/3-2b+y} - t_i^{1/3-2b+y} \right],
\end{equation}
where $\gamma_{e,i} = \gamma_e (t_i)$ and $\gamma_{e,f} = \gamma_e (t_f)$. We consider an infinitesimal group of $\delta N_e$ electrons injected at time $t_i$ during an interval [$t_i$, $t_i + \delta t_i$] above $\gamma_m$ (i.e., $\gamma_{e,i} \geq \gamma_m$). Its injection rate is given by Equation (\ref{eq:R_inj_h0}) again.
In the fast-cooling regime, these electrons cool rapidly such that $\gamma_{e,f}\, t_f^{2/3} \ll \gamma_{e,i}\, t_i^{2/3}$ is further satisfied. Equation (\ref{eq:ge_f_rad_adia}) yields then
\begin{equation}
\label{eq:ge_f_rad_adia2}
\frac{1}{\gamma_{e, f}} = 
\frac{g_0\, t_f^{2/3}}{1/3-2b+y} \left[ t_f^{1/3-2b+y} - t_i^{1/3-2b+y} \right].
\end{equation}

In order to find the electron spectrum at time $t_f$, we take a variation on both sides of Equation (\ref{eq:ge_f_rad_adia2}) at the fixed time $t_f$, which gives
\begin{equation}
\label{eq:delta_ge_f_rad_adia}
\frac{\delta \gamma_{e, f}}{[\gamma_{e, f}]^2} = 
g_0\, t_f^{2/3}\, t_i^{-2b+y-2/3}\, \delta t_i.
\end{equation}
As was in the case of radiative cooling above, Equation (\ref{eq:delta_ge_f_rad_adia}) shows that a variation in $\gamma_{e, f}$ is induced by a variation in $t_i$. This indicates that those $\delta N_e$ electrons injected at time $t_i$ during [$t_i$, $t_i + \delta t_i$] occupy an energy interval [$\gamma_{e, f}$, $\gamma_{e, f} + \delta \gamma_{e, f}$] of the electron spectrum at time $t_f$. Therefore, by eliminating $\delta t_i$ from Equations (\ref{eq:R_inj_h0}) and (\ref{eq:delta_ge_f_rad_adia}), we obtain
\begin{equation}
\label{eq:Ne_ge_f_rad_adia}
\frac{\delta N_e}{\delta \gamma_{e, f}} =
\frac{h_0}{g_0}\, t_f^{-2/3}\, [\gamma_{e, f}]^{-2}\, t_i^{2b-y+q+2/3}.
\end{equation}
Finally, solving for $t_i$ in Equation (\ref{eq:ge_f_rad_adia2}) and substituting it into Equation (\ref{eq:Ne_ge_f_rad_adia}), we find the electron spectrum
\begin{equation}
\label{eq:fU_anal_rad_adia}
\frac{\delta N_e}{\delta \gamma_e} 
= \frac{\frac{h_0}{g_0}\, t^{2b-y+q}}
{\gamma_e^2\, \left[ 1+ \left(\frac{6b-3y-1}{3g_0}\, t^{2b-y-1}\right) \frac{1}{\gamma_e} \right]^{\frac{6b-3y+3q+2}{6b-3y-1}}},
\end{equation}
where we have dropped {\it f} from $t_f$ and $\gamma_{e,f}$ for clarity. Equation (\ref{eq:fU_anal_rad_adia}) is identical to $f_{\rm U} (\gamma_e, t)$ in Equation (\ref{eq:fge_U}) when
\begin{eqnarray}
\label{eq:A_anal_rad_adia}
A(t) =&& \frac{h_0}{g_0}\, t^{2b-y+q}, \\
\label{eq:g0_anal_rad_adia}
\gamma_0(t) =&& \frac{6b-3y-1}{3g_0}\, t^{2b-y-1}, \\
\label{eq:beta_anal_rad_adia}
\beta =&& \frac{6b-3y+3q+2}{6b-3y-1}.
\end{eqnarray}
Therefore, we have analytically derived our electron spectrum $f_{\rm U}$ for radiative and adiabatic cooling of electrons.

We first note that Equation (\ref{eq:A_anal_rad_adia}) is identical to Equation (\ref{eq:A_anal_rad}): i.e., the overall magnitude $A$ remains the same between the two cooling cases. It is surprising to find that the inclusion of adiabatic cooling does not alter this expression for $A$. Secondly, the characteristic Lorentz factor $\gamma_0$ in Equation (\ref{eq:g0_anal_rad_adia}) has the same dependence on time $t$ and the cooling measure $g_0$ as in Equation (\ref{eq:g0_anal_rad}). This again suggests that the effect of adiabatic cooling is marginal in the vicinity of $\gamma_0$. On the other hand, the exponent $\beta$ in Equation (\ref{eq:beta_anal_rad_adia}) differs from Equation (\ref{eq:beta_anal_rad}): it is largely affected by the inclusion of adiabatic cooling. This finding indicates that the adiabatic cooling operates most efficiently in the low-energy part of the electron spectrum.

The ratio $h_0/g_0$ in Equation (\ref{eq:A_anal_rad_adia}) shows that the overall magnitude $A$ is regulated by a competition between the cooling and the injection. Equation (\ref{eq:g0_anal_rad_adia}) for $\gamma_0$ does not include the injection measure $h_0$ or the injection index $q$ so that it decouples from the injection. The exponent $\beta$ in Equation (\ref{eq:beta_anal_rad_adia}), which shows the low-energy behavior of the electron spectrum, is settled by the cooling and injection indices together. In a special case of $q=0$ (with a constant injection rate), it is decided by cooling only.

In this Section \ref{section:anal_deriv},
we made use of $\gamma_e \ll \gamma_m$ during the derivation. Therefore, Equations (\ref{eq:g0_anal_rad}) and (\ref{eq:g0_anal_rad_adia}) for $\gamma_0$ become less accurate if $\gamma_0 \ll \gamma_m$ is not satisfied. Equations (\ref{eq:A_anal_rad}) and (\ref{eq:A_anal_rad_adia}) for $A$ also become less accurate if $\gamma_0$ is not much smaller than $\gamma_m$. We provide a more general and accurate expression for $\gamma_0$ and $A$ in Section \ref{section:Fokker-Planck}. 

The exponent $\beta$ describes the asymptotic low-energy behavior of the electron spectrum $f_{\rm U}$. For this asymptotic part of the spectrum, $\gamma_e \ll \gamma_m$ is generally satisfied. Therefore, we emphasize that Equations (\ref{eq:beta_anal_rad}) and (\ref{eq:beta_anal_rad_adia}) for $\beta$ remain accurate even if $\gamma_0$ gets close to $\gamma_m$.

\subsection{Validity of $f_{\rm U}$ for all $\gamma_e < \gamma_m$} \label{section:Fokker-Planck}

In Section \ref{section:anal_deriv}, we assumed the power-law dependences for $B$, $Y$, and $R_{\rm inj}$ to enable our analytical derivation of $f_{\rm U}$. We also assumed $\gamma_e \ll \gamma_m$ in the middle of the derivation. Now we eliminate these assumptions and proceed below without assuming them.   

We show here that the electron spectrum $f_{\rm U}$ in Equation (\ref{eq:fge_U}), in fact, obeys the Fokker-Planck equation in Equation (\ref{eq:fokker_planck}) in the entire energy range below $\gamma_m$ (i.e., for all values of $\gamma_e < \gamma_m$).

We take a partial time-derivative on Equation (\ref{eq:fge_U}) and find
\begin{equation}
\label{eq:fU_t}
\frac{\partial f_{\rm U}}{\partial t}  =
\left( \frac{\dot A}{A} - \frac{\beta \dot \gamma_0}{\gamma_e + \gamma_0} \right) f_{\rm U},
\end{equation}
where $\dot A = {\rm d} A/ {\rm d} t$ and $\dot \gamma_0 = {\rm d} \gamma_0/ {\rm d} t$, and $\beta$ is kept constant.

\subsubsection{Radiative Cooling}

Using Equations (\ref{eq:fge_U}) and (\ref{eq:ge_dot_rad}), we get
\begin{equation}
\label{eq:ge_dot_fU_ge}
\frac{\partial (\dot \gamma_e f_{\rm U})}{\partial \gamma_e} =
-\frac{g\, \beta \gamma_0}{1+\gamma_0/\gamma_e}\, f_{\rm U}.
\end{equation}
Inserting Equations (\ref{eq:fU_t}) and (\ref{eq:ge_dot_fU_ge}) into Equation (\ref{eq:fokker_planck}) and recalling that $Q=0$ for $\gamma_e < \gamma_m$, we obtain
\begin{equation}
\label{eq:fokker_planck2_rad}
\left[ \frac{\dot A}{A} - g\, \beta \gamma_0 \right] \gamma_e +
\left[ \frac{\dot A}{A} \gamma_0 - \beta \dot \gamma_0 \right] = 0.
\end{equation}
Equation (\ref{eq:fokker_planck2_rad}) is always satisfied for all $\gamma_e$ if both brackets on the left-hand side vanish: i.e., if
\begin{equation}
\label{eq:f_p_rad_3}
\frac{\dot A}{A} - g\, \beta \gamma_0 = 0,
\end{equation}
and
\begin{equation}
\label{eq:f_p_rad_4}
\frac{\dot A}{A} \gamma_0 - \beta \dot \gamma_0 = 0.
\end{equation}
Equation (\ref{eq:f_p_rad_4}) yields
\begin{equation}
\label{eq:A_fokker_rad}
A(t) \propto \left[\gamma_0 (t) \right]^{\beta}.
\end{equation}
Combining Equations (\ref{eq:f_p_rad_3}) and (\ref{eq:f_p_rad_4}), we obtain
\begin{equation}
\label{eq:f_p_rad_5}
\dot \gamma_0  = g(t) \, \gamma_0^2.
\end{equation}
Note that $\beta$ cancels out here. Equation (\ref{eq:f_p_rad_5}) gives a general solution
\begin{equation}
\label{eq:sol_rad}
\gamma_0 (t) = \left[ C_1 - \int g(t)\, {\rm d} t \right]^{-1},
\end{equation}
where $C_1$ is an integration constant.

Therefore, $f_{\rm U}$ is a solution to the Fokker-Planck equation for all $\gamma_e$ below $\gamma_m$ for radiative cooling of electrons, if Equations (\ref{eq:f_p_rad_3}) and (\ref{eq:f_p_rad_4}) are satisfied. In other words, $f_{\rm U}$ with Equations (\ref{eq:A_fokker_rad}) and (\ref{eq:sol_rad}) provides a general electron spectrum for radiative cooling: it is valid in the entire energy range below $\gamma_m$, and it describes electrons in a non-steady state.

\subsubsection{Radiative and Adiabatic Cooling}

Using Equations (\ref{eq:fge_U}) and (\ref{eq:ge_dot_rad_adia}), we have
\begin{equation}
\label{eq:ge_dot_fU_ge2}
\frac{\partial (\dot \gamma_e f_{\rm U})}{\partial \gamma_e} =
\left[ -\frac{g\, \beta \gamma_0}{1+\gamma_0/\gamma_e} +
\frac{2}{3t} \left(1-\frac{\beta \gamma_0}{\gamma_e + \gamma_0} \right) \right] f_{\rm U}.
\end{equation}
Inserting Equations (\ref{eq:fU_t}) and (\ref{eq:ge_dot_fU_ge2}) into Equation (\ref{eq:fokker_planck}) and noting that $Q=0$ for $\gamma_e < \gamma_m$, we get
\begin{equation}
\label{eq:fokker_planck2_rad_adia}
\left[ \frac{\dot A}{A} - g\, \beta \gamma_0 +\frac{2}{3t} \right] \gamma_e +
\left[ \frac{\dot A}{A} \gamma_0 - \beta \dot \gamma_0 + \frac{2}{3t} (1-\beta) \gamma_0 \right] = 0.
\end{equation}
Equation (\ref{eq:fokker_planck2_rad_adia}) always holds for all $\gamma_e$ if both brackets on the left-hand side vanish: i.e., if 
\begin{equation}
\label{eq:f_p_rad_adia_3}
\frac{\dot A}{A} - g\, \beta \gamma_0 +\frac{2}{3t} = 0,
\end{equation}
and
\begin{equation}
\label{eq:f_p_rad_adia_4}
\frac{\dot A}{A} \gamma_0-\beta \dot \gamma_0 + \frac{2}{3t} (1-\beta) \gamma_0 = 0.
\end{equation}
Equation (\ref{eq:f_p_rad_adia_4}) gives
\begin{equation}
\label{eq:A_fokker_rad_adia}
A(t) \propto t^{2(\beta-1)/3}\, \left[\gamma_0 (t) \right]^{\beta}.
\end{equation}
Combining Equations (\ref{eq:f_p_rad_adia_3}) and (\ref{eq:f_p_rad_adia_4}), we obtain
\begin{equation}
\label{eq:bernoulli}
\dot \gamma_0 + \frac{2}{3t}\, \gamma_0 = g(t) \, \gamma_0^2,
\end{equation}
which is a Bernoulli differential equation. Note that $\beta$ cancels out here. Substituting $u \equiv 1/\gamma_0$ in Equation (\ref{eq:bernoulli}) yields
\begin{equation}
\label{eq:bernoulli2}
\frac{{\rm d} u}{{\rm d}t} - \frac{2}{3t}\, u = -g(t).
\end{equation}
Solving Equation (\ref{eq:bernoulli2}), we find
\begin{equation}
\label{eq:sol_rad_adia}
u(t) = \frac{1}{\gamma_0 (t)} = 
t^{2/3} \left[ C_2 - \int t^{-2/3}\, g(t)\, {\rm d} t \right],
\end{equation}
where $C_2$ is an integration constant.

Therefore, $f_{\rm U}$ is a solution to the Fokker-Planck equation for all $\gamma_e$ below $\gamma_m$ for radiative plus adiabatic cooling of electrons, if Equations (\ref{eq:f_p_rad_adia_3}) and (\ref{eq:f_p_rad_adia_4}) are satisfied. In other words, $f_{\rm U}$ with Equations (\ref{eq:A_fokker_rad_adia}) and (\ref{eq:sol_rad_adia}) provides a general electron spectrum for radiative plus adiabatic cooling. It is valid in the entire energy range below $\gamma_m$, and it describes electrons in a non-steady state: see Section~\ref{section:blazar} for an example.

In this Section~\ref{section:Fokker-Planck}, 
we made no assumptions, and thus the physical quantities $B$, $Y$, and $R_{\rm inj}$ remain completely general: they can be arbitrary functions of time $t$. Therefore, we highlight that the applicability of $f_{\rm U}$ is very wide and general. More examples for blazars and GRBs will be presented in an accompanying paper.

Since the function $g(t)$ in Equation (\ref{eq:g_t}) remains arbitrary, Equations (\ref{eq:sol_rad}) and (\ref{eq:sol_rad_adia}) for $\gamma_0 (t)$ provides a very general expression for the characteristic Lorentz factor (i.e., the cooling Lorentz factor). Also, since $g(t)$ is only about the cooling processes with no regard to the electron injection, Equations (\ref{eq:sol_rad}) and (\ref{eq:sol_rad_adia}) show that $\gamma_0 (t)$ is decoupled from the injection history.

We remark that the exponent $\beta$ is left undecided here: this is, in fact, a room left for the electron injection to settle. All electrons were once injected above $\gamma_m$ and cooled down to energies below $\gamma_m$. Tracking the cooling of all these electrons decides $\beta$: e.g., see the example in Section~\ref{section:blazar}. If $B$, $Y$, and $R_{\rm inj}$ follow the power-law profiles, Equations (\ref{eq:beta_anal_rad}) and (\ref{eq:beta_anal_rad_adia}) derived in Section~\ref{section:anal_deriv} still give us an accurate value for $\beta$ here.

\subsection{Consistency Check}

We analytically derived an explicit expression of $f_{\rm U}$ in Section~\ref{section:anal_deriv}. Also, using the Fokker-Planck equation, we derived a more general version of $f_{\rm U}$ in Section~\ref{section:Fokker-Planck}. We examine here whether the results in Section~\ref{section:anal_deriv} comply with those in Section~\ref{section:Fokker-Planck}.

\subsubsection{Radiative Cooling}

We first note that the analytically-derived Equations (\ref{eq:A_anal_rad}) - (\ref{eq:beta_anal_rad}) satisfy the Fokker-Planck-derived relation $A(t) \propto \left[\gamma_0 (t) \right]^{\beta}$ in Equation (\ref{eq:A_fokker_rad}).

The analytical derivation assumed the power-law dependences and used $g(t) = g_0\, t^{-2b+y}$ in Equation (\ref{eq:g_t_powerlaw}). Inserting this into the Fokker-Planck-derived solution $\gamma_0 (t)$ in Equation (\ref{eq:sol_rad}), we have
\begin{equation}
\label{eq:g0_check_rad_1}
\gamma_0 (t) = \left[ C_1 + \frac{g_0}{2b-y-1}\, t^{-(2b-y-1)} \right]^{-1}.
\end{equation}
As $t$ goes infinite, $\gamma_0$ should approach infinity since the electron cooling diminishes when $B$ vanishes: recall that $B \propto t^{-b}$. Hence, $C_1$ needs to be very small for $2b-y-1 > 0$. If $C_1 = 0$, Equation (\ref{eq:g0_check_rad_1}) yields
\begin{equation}
\label{eq:g0_check_rad_2}
\gamma_0 (t) = \frac{2b-y-1}{g_0}\, t^{2b-y-1}.
\end{equation}
Note that Equation (\ref{eq:g0_check_rad_2}) is the same as the analytically-derived Equation (\ref{eq:g0_anal_rad}).

\subsubsection{Radiative and Adiabatic Cooling}

Firstly, it is straightforward to show that the analytically-derived Equations (\ref{eq:A_anal_rad_adia}) - (\ref{eq:beta_anal_rad_adia}) satisfy the Fokker-Planck-derived relation $A(t) \propto t^{2(\beta-1)/3}\, \left[\gamma_0 (t) \right]^{\beta}$ in Equation (\ref{eq:A_fokker_rad_adia}).

The analytical derivation used $g(t) = g_0\, t^{-2b+y}$ in Equation (\ref{eq:g_t_powerlaw}). Inserting this into the Fokker-Planck-derived Equation (\ref{eq:sol_rad_adia}), we have
\begin{equation}
\label{eq:g0_check_rad_adia_1}
\gamma_0 (t) = 
t^{-2/3} \left[ C_2 + \frac{g_0}{2b-y-1/3}\, t^{-(2b-y-1/3)} \right]^{-1}.
\end{equation}
For the same reason as above, $C_2$ needs to be very small for $2b-y-1/3 > 0$. If $C_2=0$, Equation (\ref{eq:g0_check_rad_adia_1}) gives
\begin{equation}
\label{eq:g0_check_rad_adia_2}
\gamma_0 (t) = \frac{6b-3y-1}{3g_0}\, t^{2b-y-1}.
\end{equation}
Note that Equation (\ref{eq:g0_check_rad_adia_2}) is the same as the analytically-derived Equation (\ref{eq:g0_anal_rad_adia}).

\subsection{Generalization of $f_{\rm U}$ to Evolving $\beta$} \label{section:evolving_beta}

So far, we have kept the exponent $\beta$ constant. We consider here a possible evolution of $\beta$ in time. The electron spectrum $f_{\rm U}$ then has the form
\begin{equation}
\label{eq:fge_U_evolving_beta}
f_{\rm U} (\gamma_e, t)  =
\frac{A(t)}{\gamma_e^2 \left(1+\gamma_0 (t)/\gamma_e \right)^{\beta(t)}}.
\end{equation}
Taking a partial time-derivative on Equation (\ref{eq:fge_U_evolving_beta}), we find
\begin{equation}
\label{eq:fU_t_evolving_beta}
\frac{\partial f_{\rm U}}{\partial t}  =
\left( \frac{\dot A}{A} - \frac{\beta \dot \gamma_0}{\gamma_e + \gamma_0} - 
\dot \beta\, \ln \left(1+\frac{\gamma_0}{\gamma_e} \right) \right) f_{\rm U},
\end{equation}
where $\dot \beta = {\rm d} \beta/ {\rm d} t$.

When compared to Equation (\ref{eq:fU_t}), we have an additional term on the right-hand side, which involves $\dot \beta$. This additional term vanishes for $\gamma_e \gg \gamma_0$ regardless of the value of $\dot \beta$. Therefore, even when $\beta$ evolves in time, $f_{\rm U}$ in Equation (\ref{eq:fge_U_evolving_beta}) still obeys the Fokker-Planck equation in the energy range $\gamma_e \gg \gamma_0$. Also, since the logarithmic factor $\ln (1+\gamma_0/\gamma_e)$ is only a few even for $\gamma_e \ll \gamma_0$, this additional term can be negligibly small compared to the first and second terms for small values of $\dot \beta$. This suggests that, for slowly varying $\beta$, $f_{\rm U}$ in Equation (\ref{eq:fge_U_evolving_beta}) still closely follows the Fokker-Planck equation and provides a general electron spectrum in the entire energy range below $\gamma_m$: see the example in Section~\ref{section:blazar}.

Equations (\ref{eq:sol_rad}) and (\ref{eq:sol_rad_adia}) for $\gamma_0 (t)$ in Section \ref{section:Fokker-Planck} remain accurate even when $\beta$ evolves in time: this is because $\beta$ canceled out from the differential Equations (\ref{eq:f_p_rad_5}) and (\ref{eq:bernoulli}), respectively.

On the other hand, if $\beta$ evolves in time, Equations (\ref{eq:A_fokker_rad}) and (\ref{eq:A_fokker_rad_adia}) for $A(t)$ in Section \ref{section:Fokker-Planck} become less accurate since the step from Equation (\ref{eq:f_p_rad_4}) to (\ref{eq:A_fokker_rad}) and the step from Equation (\ref{eq:f_p_rad_adia_4}) to (\ref{eq:A_fokker_rad_adia}) made use of $\dot \beta = 0$. 

We clarify that Equations (\ref{eq:A_fokker_rad}) and (\ref{eq:A_fokker_rad_adia}) being less accurate due to evolving $\beta$ does not mean that $f_{\rm U}$ in Equation (\ref{eq:fge_U_evolving_beta}) fails to obey the Fokker-Planck equation. The spectrum $f_{\rm U}$ in Equation (\ref{eq:fge_U_evolving_beta}) still obeys the Fokker-Planck equation when Equations (\ref{eq:f_p_rad_4}) and (\ref{eq:f_p_rad_adia_4}) are satisfied.

%% ----------------------------------
%% The asterisk after "figure" tells the compiler to span multiple columns if a two column style is selected.

\begin{figure}[t!]
\plotone{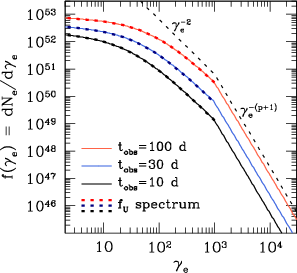}
\caption{
Comparison of our electron spectrum $f_{\rm U}$ (thick-dotted lines) to a detailed numerical calculation of a blazar electron spectrum (solid lines) as a function of observer time $t_{\rm obs}$. The solid lines show the model [a] presented in \cite{uhm2026a}. The model [a] has a minimum injection Lorentz factor $\gamma_m = 10^3$.
\label{fig:f1}}
\end{figure}
%% ----------------------------------

%% ----------------------------------
\begin{deluxetable*}{lcccc}
\tabletypesize{\scriptsize}
\tablewidth{0pt} 
\tablecaption{Parameters of the electron spectrum $f_{\rm U}$ plotted in Figure~\ref{fig:f1} for observer times 10 d, 30 d, and 100 d \label{tab:fU_parameters}}
\tablehead{
\colhead{} & \colhead{} & \multicolumn{3}{c}{Observer time $t_{\rm obs}$} \\
\cline{3-5}
\colhead{Parameters} & \colhead{} & \colhead{10 d} & \colhead{30 d} & \colhead{100 d} 
} 
%\colnumbers
\startdata 
$A$         & &  $1.4 \times 10^{55}$  &  $6.5 \times 10^{55}$  &  $3.7 \times 10^{56}$ \\
$\gamma_0$  & &  30                    &  47.5                  &  79                   \\
$\beta$     & &  1.91                  &  1.92                  &  1.93                 \\
\enddata
\tablecomments{Our electron spectrum $f_{\rm U}$ in Equation (\ref{eq:fge_U}) takes the form, $f_{\rm U} (\gamma_e) = A\, \gamma_e^{-2}\, \left(1 + \gamma_0/\gamma_e \right)^{-\beta}$.}
\end{deluxetable*}
%% ----------------------------------

\section{Application to Blazars} \label{section:blazar}

Here, we apply our electron spectrum $f_{\rm U}$ to blazars. We compare $f_{\rm U}$ to a detailed numerical calculation of a blazar electron spectrum presented in \cite{uhm2026a}.

\cite{uhm2026a} computes the blazar electron spectrum numerically by employing a sophisticated method, in which the continuously injected electrons are divided into infinitesimally small divisions in the electron energy space and also in time space. Then, the cooling of each infinitesimal group of electrons is individually tracked while taking into account the radiative and adiabatic cooling of electrons. The electron injection occurs only above $\gamma_m$ and has a power-law distribution $\gamma_e^{-p}$.

Figure \ref{fig:f1} shows a comparison of $f_{\rm U}$ to the blazar electron spectrum of the model [a] presented in \cite{uhm2026a}: the thick-dotted lines show the $f_{\rm U}$ spectrum, and the solid lines show the blazar electron spectrum at three observer times $t_{\rm obs} =$ 10, 30, and 100 days. As one can see, the $f_{\rm U}$ spectrum is nearly identical to the blazar electron spectrum: we stress that we have an excellent agreement in the entire energy range below $\gamma_m = 10^3$. The $A$, $\gamma_0$, and $\beta$ values of the $f_{\rm U}$ spectrum plotted in Figure \ref{fig:f1} are summarized in Table~\ref{tab:fU_parameters}. 

Table~\ref{tab:fU_parameters} shows that $\beta$ slowly evolves in time. Hence, we highlight that, even when $\beta$ evolves in time, $f_{\rm U}$ still provides a general electron spectrum to describe the entire cooled population of electrons below $\gamma_m$. We remark that the electron spectrum in Figure \ref{fig:f1} evolves over time and thus those electrons are indeed in a non-steady state.

\section{Conclusion and Discussion}

We presented a new electron spectrum $f_{\rm U}$ produced by cooling of relativistic electrons in the fast-cooling regime. We showed that the spectrum $f_{\rm U}$ gives us a general electron spectrum to describe the entire cooled population of electrons below $\gamma_m$, which is in a non-steady state. We showed that $f_{\rm U}$ is a solution to the Fokker-Planck equation below $\gamma_m$ and gives us a new fast-cooling solution below $\gamma_m$. We showed that $f_{\rm U}$ provides a general spectrum below $\gamma_m$ for two cooling cases: (1) when the electrons undergo radiative (i.e., synchrotron + IC) cooling, and (2) when the electrons experience both radiative and adiabatic cooling. 

We first derived $f_{\rm U}$ analytically and found an explicit expression of $f_{\rm U}$ for each cooling case. Secondly, we showed that $f_{\rm U}$ is a solution to the Fokker-Planck equation in the entire energy range below $\gamma_m$ (i.e., for all $\gamma_e < \gamma_m$) and derived a general version of $f_{\rm U}$ for each cooling case. We showed that the results of the first analytical derivation comply with those of the second Fokker-Planck derivation.

Our electron spectrum $f_{\rm U}$ consists of three constituents $A(t)$, $\gamma_0(t)$, and $\beta$. We found that the overall magnitude $A(t)$ is regulated by a competition between the electron cooling and injection, and the characteristic Lorentz factor $\gamma_0(t)$ is decoupled from the electron injection and physically corresponds to a cooling Lorentz factor, which is usually referred to as $\gamma_c$ in the field. The exponent $\beta$, which shapes the asymptotic low-energy part of the spectrum, is influenced by the cooling and injection together. In a special case with a constant injection rate, the exponent $\beta$ is decided by cooling only.

We showed that the spectrum $f_{\rm U}$ can be generalized to include evolving $\beta$. For slowly varying $\beta$, $f_{\rm U}$ still satisfies the Fokker-Planck equation and gives us a general electron spectrum for the entire energy range below $\gamma_m$. 

We applied $f_{\rm U}$ to blazars. We compared $f_{\rm U}$ to a detailed numerical calculation of a blazar electron spectrum and showed that $f_{\rm U}$ is nearly identical to the blazar electron spectrum for all $\gamma_e < \gamma_m$ even when $\beta$ evolves in time. In other words, $f_{\rm U}$ indeed describes the entire cooled population of electrons below $\gamma_m$ while those electrons evolve in a non-steady state.

The electron spectrum $f_{\rm U}$ does not contain information about the injected electron spectrum above $\gamma_m$ such as a power-law index $p$, which indicates that the ``cooling physics'' governing the energy loss of electrons completely overwrites this information of injected spectrum. Information about the injection rate of electrons is not lost and affects $f_{\rm U}$ through $A(t)$ and $\beta$.

By taking $y=0$ and $Y_0=1$ in the analytical derivation with the power-laws, or by setting $Y=0$ in the Fokker-Planck derivation, we can obtain another expressions of $f_{\rm U}$ for the case of no IC cooling (i.e., the radiative cooling includes synchrotron cooling only). 

For instance, by setting $y=0$ and $Y_0=1$ in Equation (\ref{eq:fU_anal_rad}), we find the following spectrum for the case where the electrons undergo synchrotron cooling:
\begin{equation}
\label{eq:fU_anal_syn}
f_{\rm U} (\gamma_e, t) 
= \frac{\frac{h_0}{g_0}\, t^{2b+q}}
{\gamma_e^2\, \left[ 1+ \left(\frac{2b-1}{g_0}\, t^{2b-1}\right) \frac{1}{\gamma_e} \right]^{\frac{2b+q}{2b-1}}}.
\end{equation}
Also, by letting $y=0$ and $Y_0=1$ in Equation (\ref{eq:fU_anal_rad_adia}), we obtain the following spectrum for the case where the electrons experience both synchrotron and adiabatic cooling:
\begin{equation}
\label{eq:fU_anal_syn_adia}
f_{\rm U} (\gamma_e, t) 
= \frac{\frac{h_0}{g_0}\, t^{2b+q}}
{\gamma_e^2\, \left[ 1+ \left(\frac{6b-1}{3g_0}\, t^{2b-1}\right) \frac{1}{\gamma_e} \right]^{\frac{6b+3q+2}{6b-1}}}.
\end{equation}
Note that $g_0$ in Equations (\ref{eq:fU_anal_syn}) and (\ref{eq:fU_anal_syn_adia}) now reads
\begin{equation}
g_0 \equiv \frac{\sigma_T}{6 \pi m_e c}\, B_0^2\, (c \Gamma/r_0)^{-2b}.
\end{equation}
Our electron spectrum in Equations (\ref{eq:fU_anal_syn}) and (\ref{eq:fU_anal_syn_adia}) are applicable in deep Klein-Nishina regime where the IC scatterings are strongly suppressed.

In this work, we took the IC cooling into account in the Thomson regime, in which the IC cooling rate of electrons has a simple dependence on the electron energy as $\dot \gamma_e \propto -\gamma_e^2$, just like for the synchrotron cooling. This simplicity, in fact, enabled us to derive our electron spectrum $f_{\rm U}$ while keeping the IC cooling included in our calculations. In Klein-Nishina (KN) regime, the KN cross-section of the IC scatterings has a complicated dependence on the electron energy, and so does the IC cooling rate of electrons \citep{nakar2009,barniol_duran2012,aguilar-ruiz2026}. This fact hinders similar derivations as presented here, thus limiting the usage of our electron spectrum $f_{\rm U}$ in moderate KN regime.

As noted in Equation (\ref{eq:fge_U_asym}), the spectrum $f_{\rm U}$ has an asymptotic slope -2 well above the cooling Lorentz factor $\gamma_0$. If $\gamma_0$ is sufficiently smaller than $\gamma_m$, this asymptotic slope -2 can show up in the electron spectrum in a region well above $\gamma_0$ but below $\gamma_m$. In other words, $f_{\rm U}$ can show the widely-known index -2 in a region near $\gamma_m$ when the cooling is sufficiently fast. However, even in this case, the electron spectral slope begins to differ from and gets significantly deviated from -2 as one move away from this region of slope -2 toward the lower energies. The electron slope approaches an asymptotic value of $-2+\beta$ at $\gamma_e \ll \gamma_0$.

The low-energy hump of the blazar spectrum is usually explained by synchrotron emission from relativistic electrons \citep{marscher1985,bottcher2007}, and its spectral shape is well described by a ``log-parabola'' function \citep{massaro2004,cheong2024,zuo2025}. \cite{uhm2026a} recently showed that the synchrotron emission from the cooled population of electrons below $\gamma_m$ in the blazar electron spectrum produces this log-parabola spectrum. Thus, an observational signature of our electron spectrum $f_{\rm U}$ is actually a log-parabola function: Details will be presented in the accompanying paper.

This work considers a single region of electrons in which freshly accelerated electrons are continuously injected. In an environment of astrophysical shocks like in GRBs and blazars, the electrons accelerated at the shock front are advected into the downstream, and the downstream region is often divided into numerous infinitesimal fluid elements or shells so that the physical processes in each shell can be individually traced \citep{granot2002,uhm2012,uhm2014b,ferguson2026}. A comparison and extension of the current work to such divided downstreams will be investigated in a future work.

We highlight that our finding of a new fast-cooling electron spectrum below $\gamma_m$ may find it applicable in various energetic sources in high-energy astrophysics, such as GRBs, blazars, pulsar wind nebulae, supernova remnants, tidal disruption events, etc.

%% Please use the acknowledgment and contribution environments. This will 
%% be anonomyized when the "anonymous" style option is used. 

\begin{acknowledgments}
We thank the anonymous referee for many helpful and constructive comments.
We also thank Andrei M. Beloborodov for helpful comments. This work was supported by the main research project (No. 2026183200) from Korea Astronomy and Space Science Institute (KASI).
\end{acknowledgments}


%% For this sample we use BibTeX plus aasjournalv7.bst to generate the

%\bibliography{ms}{}
%\bibliographystyle{aasjournalv7}



\end{document}